\documentclass{iopart}
\usepackage{iopams,bm,cite,color,graphicx,subfigure,braket,mathptmx,bm,color,url}

\eqnobysec

\usepackage{graphicx,array,enumitem,epstopdf,boxedminipage,bm,color}

\newcommand{\sugg}[1]{{#1}}
\usepackage[colorlinks=true,linkcolor=blue,citecolor=blue,urlcolor =blue]{hyperref}
\newcolumntype{L}{>{$}l<{$}} 
\newcommand{\opinner}[3]{\left<{#1}\vphantom{#1}\vphantom{#3}\right|{#2}\left|{#3}\vphantom{#1}\vphantom{#3}\right>}

\newcommand{\Span}[1]{\mathrm{span}\!\left\{#1\right\}}
\newcommand{\Hsub}{\mathcal{H}_\mathrm{sub}}
\newcommand{\Hphys}{\mathcal{H}_\mathrm{phys}}
\newcommand{\dphys}{d_\mathrm{phys}}
\newcommand{\RE}{\mathop{\mathrm{Re}} \nolimits}
\newcommand{\IM}{\mathop{\mathrm{Im}} \nolimits}
\newcommand{\MEAN}[2]{\mathbb{E}_{#1}\!\left[#2\right]}
\renewcommand{\tr}[1]{\mathrm{tr} ( #1 )}

\begin{document}
	
\title{Extracting the physical sector of quantum states}
\author{D~Mogilevtsev$^{1,2}$, Y~S~Teo$^3$, J~{\v R}eh{\'a}{\v c}ek$^4$, 
Z~Hradil$^4$, J~Tiedau$^5$, R~Kruse$^5$, G~Harder$^5$, C~Silberhorn$^{5,6}$, 
L~L~Sanchez-Soto$^{5,6}$}

\address{$^{1}$ Centro de Ci{\^e}ncias Naturais e Humanas, Universidade Federal do ABC, 
 Santo Andr{\'e}, SP, 09210-170 Brazil}
\address{$^{2}$   B. I. Stepanov Institute of Physics, National
  Academy of Science of Belarus,  Nezavisimosti Avenue 68, 
Minsk 220072, Belarus}
\address{$^{3}$ BK21 Frontier Physics Research Division, 
Seoul National University, 08826 Seoul, South Korea}
\address{$^{4}$ Department of Optics, Palack{\'y} University, 
17. listopadu 12, 77146 Olomouc, Czech Republic}
\address{$^{5}$ Department Physik, Universit\"{a}t Paderborn,
	Warburger Stra{\ss}e~100, 33098 Paderborn, Germany}
\address{$^{6}$ Max-Planck-Institut f\"ur  die Physik des Lichts,
	Staudtstra\ss e 2, 91058 Erlangen, Germany}
\address{$^{7}$ Departamento de \'Optica, Facultad de F\'{\i}sica,
	Universidad Complutense, 28040 Madrid, Spain}

      \begin{abstract}
        The physical nature of any quantum source guarantees the
        existence of an effective Hilbert space of finite dimension,
        the physical sector, in which its state is completely
        characterized with arbitrarily high accuracy. The extraction
        of this sector is essential for state tomography. We show that
        the physical sector of a state, defined in some pre-chosen
        basis, can be systematically retrieved with a procedure using
        only data collected from a set of commuting quantum
        measurement outcomes, with no other assumptions about the
        source. We demonstrate the versatility and efficiency of the
        physical-sector extraction by applying it to simulated and
        experimental data for quantum light sources, as well as
        quantum systems of finite dimensions.
      \end{abstract}
       \submitto{\NJP}

\section{Introduction}

The physical laws of quantum mechanics ensure that all experimental
observations can be described in an \emph{effective} Hilbert space of
finite dimension, to which we shall refer as the \emph{physical
  sector} of the state. The systematic extraction of this
physical sector is crucial for reliable quantum state tomography.

Photonic sources constitute an archetypical example where such an
extraction is indispensable. Theoretically, the states
describing these sources reside in an infinite-dimensional Hilbert
space. Nonetheless, the elements of the associated density matrices
decay to zero for sufficiently large photon numbers, so that there
always exists a finite-dimensional physical sector that contains the
state with sufficient accuracy. Reliable state tomography can thus be
performed once this physical sector is correctly extracted.

Experiments on estimates of the correct physical sector have been
carried out~\cite{lnp:2004uq,Lvovsky:2009ys}.  \sugg{ One common
  strategy is to make an educated guess about the state (such as
  Gaussianity~\cite{Rehacek:2009aa} or rank-deficiency for compressed
  sensing~\cite{Gross:2010dq,Cramer:2010oq,Flammia:2012if,
    Landon-Cardinal:2012mb,Baumgratz:2013fq,Riofrio:2017aa,Steffens:2017aa}),
  which defines a truncated reconstruction subspace. For instance, in
  compressed sensing the rank of the state is assumed to be no larger
  than a certain value $r$, so that specialized rank-$r$
  compressed-sensing measurements can be employed to uniquely
  characterize the state with much fewer measurement settings. Very
  generally, educated guesses of certain properties of the state
  requires additional physical verifications. Algorithms for
  statistical model selection, such as the Akaike~\cite{Akaike:1974af,
    Usami:2003aa,Yin:2011aa} or Schwarz criteria~\cite{Schwarz:1978aa,
    Guta:2012bl} or the ìlikelihood
  sieve~\cite{Geman:1982aa,Artiles:2005aa}, have also been developed
  to estimate the physical sector. These algorithms provide another
  practical solution to reducing the complexity of the tomography
  problem. In the presence of the positivity
  constraint~\cite{Anraku:1999aa,Hughes:2003aa}, their application to
  quantum states becomes more sophisticated, as the procedures for
  deriving stopping criteria that supplies the final appropriate model
  subspace for the unknown state are intricate.}

On the other hand, finite-dimensional systems represent
another example for which a systematic physical-sector extraction
becomes important. In the context of quantum information, ongoing
developments in dimension-witness testing~\cite{Brunner:2008aa,
Hendrych:2012aa,Ahrens:2012aa,Brunner:2013aa,Ahrens:2014aa}
offer some solutions to finding the minimal dimension of a 
black box required to justify the given set of measurement data in a
device-independent way. Searching for dimension witnesses of arbitrary
dimensions is still challenging~\cite{Brunner:2013aa}.

In reference~\cite{Rehacek:2016ch}, we showed that, when the
measurement device is calibrated, one can systematically extract the
physical sector (that is, both the Hilbert-space support and
dimension) and simultaneously reconstruct any unknown state
directly from the measurement data without any assumption about the
state. In this paper, we introduce an even more efficient procedure
that extracts the physical sector of any state from the data without
state reconstruction and provide the pseudocode. This procedure
requires nothing more than data obtained from a set of commuting
measurements. As in~\cite{Rehacek:2016ch}, the extraction of the
physical sector does not depend on any other assumptions or
calibration details about the source. By construction, this procedure
has a linear complexity in the dimension of the physical sector. To
showcase its versatility, we apply it to simulated and experimental
data for photonic sources and  systems of finite dimensions. In
this way, we offer a deterministic solution to the problem of
extracting the correct physical sector for any quantum state in
measurement-calibrated situations.

\section{Physical sectors and commuting measurements}

\subsection{What are physical sectors?}

The concept of physical sectors and their relations to commuting
measurements is probably best understood with a concrete example. Let us
consider, in the Fock basis, a quantum state of light described by the
density operator
\begin{equation}
\varrho \,\,\widehat{=}
\pmatrix{
	0.9922 & * & 0.0877 & * &  \cdots \cr
	* & * & * & * & \cdots\cr
	0.0877 & * & 0.0078 & * &\cdots \cr
	* & * & * & * &\cdots \cr
	\vdots & \vdots & \vdots & \vdots &\ddots
} \,,
\label{eq:rho_EX}
\end{equation}
where $*$ denotes elements of its density matrix that are so tiny that
treating them to be zero incurs very small truncation errors. If all
$*=0$, $\varrho$ is the pure state $\ket{\,\,\,}\bra{\,\,\,}$ described
by $\ket{\,\,\,}\propto\ket{\alpha}+\ket{-\alpha}$, with the coherent
state of amplitude $\alpha=0.3536$. The density matrix elements drops
to zero for sufficiently large photon numbers as those of any physical
state.

Some statistical reasoning for understanding the truncation error is
in order. For now, we note that since all other $*$ elements are tiny,
the state $\varrho$ is essentially fully characterized by a
3-dimensional sector, such that elements beyond this sector supply
almost no contribution to $\varrho$. This forms a truncated Hilbert
subspace where tomography can be carried out reliably. This subspace
is given by $\Hsub=\Span{\ket{0},\ket{1},\ket{2}}$. However from
\eref{eq:rho_EX}, we realize that this subspace is not the smallest
one that supports $\varrho$. The \emph{smallest} subspace
$\Hphys=\Span{\ket{0},\ket{2}}$ is in fact spanned by only two basis
kets. This defines the 2-dimensional \emph{physical sector}.

In general, the physical sector $\Hphys$ is defined to be the
\emph{smallest Hilbert subspace} that fully supports a given 
state with a truncation error smaller than some tiny $\epsilon$ in
some basis. Evidently, the choice of basis affects the description of
$\Hphys$. If one already knows that $\varrho$ is close to
$\ket{\,\,\,}\bra{\,\,\,}$, then choosing $\ket{\,\,\,}$ as part of a
basis gives a 1-dimensional $\Hphys$. Such knowledge is of course
absent when $\varrho$ is unknown. In such a practical scenario in
quantum optics, we may adopt the most common Fock basis for
representing $\varrho$ and $\Hphys$. When dealing with general quantum
systems, the basis that is most natural in typical experiments may be
chosen, such as the Pauli computational basis for qubit systems.

\subsection{How are physical sectors related to commuting measurements?}

Let us revisit the example in \eref{eq:rho_EX}. Because of the 
positivity constraint imposed on $\varrho$, whenever a diagonal
element is $*$, then elements in the row and column that intersect
this element are all $*$. Also, if a diagonal element is not $*$, then
it is obvious that $\Hphys$ is spanned by the basis ket for this
diagonal element. For this example, the 2-dimensional $\Hphys$
completely characterizes $\varrho$ with the $2^2=4$ elements
$\varrho_{00}$, $\varrho_{22}$, $\RE (\varrho_{02} )$ and
$\IM ( \varrho_{02})$.

It follows that knowing the location of \emph{significant}
diagonal elements are all we need to ascertain $\Hphys$. For this
purpose the only necessary tool is a set of \emph{commuting}
measurement outcomes with their common eigenbasis being the pre-chosen
basis for $\Hphys$. After the measurement data are performed with
these commuting outcomes, all one needs to do is perform an extraction
procedure on the data to obtain $\Hphys$. This procedure would proceed
to test a growing set of basis kets until it informs that the current
set spans $\Hphys$ that fully supports the data. We note here that the
extraction works for any other sort of generalized measurements in
principle, although we shall consider commuting measurements in
subsequent discussions since they are the simplest kind
necessary for extracting physical sectors in large Hilbert-space
dimensions.

\section{The extraction of the physical sector}

In some pre-chosen basis, the physical-sector extraction procedure
(PSEP) iteratively checks whether its data are supported by the
cumulative sequence of $\Hsub$ with truncation error smaller than some
tiny $\epsilon$. PSEP starts deciding whether, say,
$\Hsub=\Span{\ket{n_1},\ket{n_2}}$ of the smallest dimension $d=2$
adequately supports the data. If yes, it takes this as the
2-dimensional $\Hphys$. Otherwise, PSEP continues and decides if
$\Hsub=\Span{\ket{n_1},\ket{n_2},\ket{n_3}}$ adequately supports the
data, and so on until finally PSEP assigns a $\dphys$-dimensional
$\Hsub=\Hphys$ with some statistical reliability. In each iterative
step, there are three objectives to be met:
\begin{enumerate}[label=\textbf{(C\roman*)}]
\item\label{pt:decide}
PSEP must decide if the data are supported with
  $\Hsub$ spanned by some set of basis kets or
  not.  
\item\label{pt:error}
PSEP must report the reliability of the
  statement ``$\Hsub$ supports $\varrho$ with truncation error less
  than $\epsilon$''.
\item\label{pt:smallest} 
PSEP must ensure that the final accepted set
  of basis kets span $\Hphys$, the \emph{smallest} $\Hsub$ that
  supports $\varrho$.
\end{enumerate}
In what follows, we show that all these objectives can be fulfilled
with only the information encoded in the measurement data.

\subsection{Deciding whether the data are supported with some subspace}

We proceed by first listing a few notations. In an experiment, a set
of measured commuting outcomes are described by positive operators
$\sum_j \Pi_j=1$. They give measurement probabilities
$p_j=\tr{\varrho \Pi_j}$ according to the Born rule. Each
commuting outcome, in the common eigenstates $\ket{n}\bra{n}$ that are
also used to represent the physical sector, can be written as
\begin{equation}
  \Pi_j=\sum_l c_{jl}\ket{l} \bra{l}
  \label{eq:diagp}
\end{equation}
with positive weights $c_{jl}$ that characterize the outcome.

To decide whether the $p_j$s are supported with some Hilbert subspace
$\Hsub$, the easiest way is to introduce Hermitian \emph{decision
  observables}
\begin{equation}
  W_\mathrm{sub}=\sum_jy_j\Pi_j
\end{equation}
for real parameters $y_j$. The decision observable for testing
$\Hsub$, along with its $y_j$s, satisfies the defining property,
\begin{equation}
	\opinner{n}{W_\mathrm{sub}}{n}=\left\{\begin{array}{@{\kern2.5pt}lL}
	\hfill 0 & if $\ket{n}\in\Hsub$\,,\\
	\hfill a_n>0 & otherwise\,.
	\end{array}\right.
	\label{eq:dec_obs}
\end{equation}
This property automatically ensures that if $\varrho$ is
\emph{completely} supported in $\Hsub$, then the expectation value
$\left<W_\mathrm{sub}\right>=\sum_jy_jp_j=0$ with zero truncation
error and PSEP takes this to be the physical sector
($\Hsub=\Hphys$). Quantum systems of finite dimensions possess states
of this kind. In quantum optics however, $\varrho$ is not completely
supported in any subspace, but possesses decaying density-matrix
elements with increasing photon numbers [such as the example in
\eref{eq:rho_EX}]. A laser source, for instance, cannot produce light
of an infinite intensity. Furthermore, the Born probabilities $p_j$
are never measured. Instead, the data consist of relative frequencies
$f_j$ that estimate the probabilities with statistical
fluctuation. Therefore, if we define the decision random variable (RV)
\begin{equation}
  w_\mathrm{sub}=\sum_jy_jf_j
\end{equation}
that estimates $\left<W_\mathrm{sub}\right>$, then PSEP may assign
$\Hsub=\Hphys$ with a truncation error defined by $|w_\mathrm{sub}|$
that is smaller than $\epsilon$.

\subsection{Quantifying the reliability of the truncation error report}

The decision RV $w_\mathrm{sub}$ is an unbiased RV in that the data
average of $w_\mathrm{sub}$ is the true value
$\left<W_\mathrm{sub}\right>$ that PSEP achieves to estimate
($\MEAN{}{w_\mathrm{sub}}=\left<W_\mathrm{sub}\right>$). This means
that in the limit of large number of measured detection events $N$ for
the data $\{f_j\}$, $w_\mathrm{sub}$ approaches its expected value
$\MEAN{}{w_\mathrm{sub}}$, which in turn tends to zero in the limit
$\Hsub\rightarrow\Hphys$. This limiting behavior invites us to
understand the truncation error $|w_\mathrm{sub}|$ using the
well-known Hoeffding inequality~\cite{Hoeffding:1963aa}, which states
that
\begin{equation}
  \alpha\equiv{\rm  Pr}
  \left \{
    |w_\mathrm{sub}| \geq \epsilon
  \right \} \leq
  2\exp\!\left(-\frac{N\epsilon^2}{2 \sum_jy_{j}^2}\right)\,.
  \label{eq:bound}
\end{equation}
This concentration inequality directly bounds the probability $\alpha$
of having a truncation error greater than or equal to $\epsilon$,
which is the significance level of the hypothesis that
$w_\mathrm{sub}=\MEAN{}{w_\mathrm{sub}}$ for all conceivable future
data~\cite{Sampling:2014aa}. With
\begin{equation}
  N\geq-\frac{2\ln(\alpha/2)}{\epsilon^2}\sum_jy_{j}^{2}\,,
\end{equation}
we are assured with $\alpha$ significance that the main factor for a
nonzero $|w_\mathrm{sub}|$ comes from insufficient support from
$\Hsub$ since statistical fluctuation is heavily suppressed.

One can obtain the more experimentally-friendly
inequality~\cite{Hoeffding:1963aa}
\begin{equation}
  \alpha\leq B_\mathrm{sub}=2\exp\!\left(-\frac{|w_\mathrm{sub}|^2}
{2\Delta^2}\right)
  \label{eq:crit}
\end{equation}
in terms of the variance $\Delta^2$ of $w_\mathrm{sub}$, where we take
$\epsilon\approx|w_\mathrm{sub}|$ as a sensible guide to the
truncation-error threshold. For $N\gg1$, the $1/N$ scaling of
$\Delta^2$ allows the quantity $B_\mathrm{sub}$ to provide an
indication on the reliability of the statement ``$\Hsub$ supports
$\varrho$ with truncation error less than $\epsilon$'' with a
reasonable statistical estimate for $\Delta^2$ from the data. If
\eref{eq:crit} holds for $\Hsub$ and some pre-chosen $\alpha$, then
the assignment $\Hphys=\Hsub$ is made. Quite generally,
$w_\mathrm{sub}$ and $\Delta^2$ reveal the influence of both
statistical and systematic
errors~\cite{Mogilevtsev:2013aa}. Therefore, by construction, for
sufficiently large $N$, $\Hsub$ eventually converges to the unique
$\Hphys$ at $\alpha$ significance with increasing size of the basis
set for properly chosen $\Hsub$. The choice of $\Hsub$ at each
iterative step of PSEP must be made so that the final extracted
support is indeed $\Hphys$, the smallest support for $\varrho$.

\subsection{Ensuring that the physical sector is extracted, not another larger support}
\label{subsec:Hphys_not_l}

To ensure that $\Hphys$ is really extracted, and not
some other larger $\Hsub$ that also supports the data, we once more
return to the example in \eref{eq:rho_EX}. For that pure state, in the
Fock basis, the $\Hsub$ that supports the state is effectively
3-dimensional, whereas $\Hphys$ is effectively 2 dimensional. With
sufficiently large number of detection events $N$, if one naively
carries out PSEP starting from $\Hsub=\Span{\ket{0}}$, PSEP would
recognize that $\Hsub$ cannot support the data, continue to test the
next larger subspace $\Hsub=\Span{\ket{0},\ket{1}}$, where it would
again conclude insufficient support. Only after the third step will
PSEP accept $\Hsub=\Span{\ket{0},\ket{1},\ket{2}}$ as the support at
some fixed $\alpha$ significance. However, $\Hsub\neq\Hphys$.

In order to efficiently extract $\Hphys$, we need only one additional
clue \emph{from the data}, that is the relative size of the diagonal
elements of $\varrho$. We emphasize here that we are \emph{not}
interested in the precise values of the diagonal elements, but only a
very rough estimate of their relative ratios to guide PSEP. With this
clue, we can then apply PSEP using the appropriately ordered sequence
of basis kets to most efficiently terminate PSEP and obtain the
smallest possible support for the data. For the pure-state example,
the decreasing magnitude of the diagonal elements gives the order
$\{\ket{0},\ket{2}\}$. For any arbitrary set of commuting $\Pi_j$s,
given the measurement matrix $\bm{C}$ of coefficients $c_{jl}$,
sorting the column $\bm{C}^{-}\bm{f}$, defined by the Moore-Penrose
pseudoinverse $\bm{C}^{-}$ of $\bm{C}$, in descending order suffices
to guide PSEP\footnote{This is \emph{not} tomography for the photon
  number distribution, but merely a very rough estimate on the
  relative ratios of diagonal elements, since $\bm{C}^{-}\bm{f}$ is
  not positive.}. This sorting permits the efficient completion of
PSEP in $O(\dphys)$ steps without doing quantum tomography. Other
sorting algorithms are, of course, possible without any information
about the diagonal-element estimates. One can perform other tests on
different permutations of basis kets within the extracted
Hilbert-subspace support, although the number of steps required to
complete PSEP would be larger than $O(\dphys)$.

\subsection{An important afterword on physical-sector extraction}

An astute reader would have already noticed that it is the $\Hphys$
\emph{within} the field-of-view (FOV) of the data that can be reliably
extracted. The FOV is affected by three factors: the degree of linear
independence of the measured outcomes, the choice of some very large
subspace to apply PSEP whose dimension does not exceed this degree of
linear independence, and the accuracy of the data (the value of
$N$). In real experiments, the number of linearly independent outcomes
measured is always finite. With the corresponding finite data set,
there exists a large subspace for extracting $\Hphys$, in which the
decision observables $W_{\mathrm{sub}}$ always satisfy \eref{eq:dec_obs} for
any $\Hsub$. For sufficiently large $N$, the collected data will
capture all significant features of $\Hphys$ within this data FOV.

Indeed, if the source is \emph{truly} a black box, then defining the
data FOV can be tricky. True black boxes are, however, atypical in a
practical tomography experiment since it is usually the observer who
prepares the state of the source and can therefore be confident that
the state prepared should not deviate too far from the target state as
long as the setup is reasonably well-controlled. The data FOV should
therefore be guided by this common sense. On the other hand, the
extraction of $\Hphys$ in device-independent cryptography, where both
the source and measurement are completely untrusted for arbitrary
quantum systems, is still an open problem.

We note here that the measurement in \eref{eq:diagp}
may incorporate realistic imperfections, such as noise,
finite detection efficiency, that are faced in a number of realistic schemes. 
For instance, the commuting diagonal outcomes may represent on/off detectors of varying efficiencies, or incorporate thermal noise \cite{prl2006,prl2014}. All such measurements are presumed to be calibratable, as non-calibrated measurements require other methods to probe the source. As an example, suppose that the measurement is inefficient but still trustworthy enough for the observer to describe its outcomes by the set $\{\eta_j\Pi_j\}$ with unknown inefficiencies $\eta_j<1$ that are simple functions of a few practical parameters of the setup such as transmissivities, losses and so forth.  In other words, we have $\eta_j=\eta_j(T_1,\ldots,T_l)$ for $l$ that is typically much less than the total number of outcomes in practical experiments. Then the straightforward practice is to first calibrate all $T_j$s before using them to subsequently carry out PSEP for other sources. One may also choose to calibrate $T_j$ already during the sorting stage by ``solving'' the linear system $\bm{t}=\bm{C}^{-}\bm{f}'$, where $f'_j=f_j/\eta_j$ is now linear in the data $f_j$ and nonlinear in $T_j$. The estimation of $T_j$ falls under parameter tomography that is beyond the scope of this discussion, which focuses on the idea of locating physical sectors and not the exact values of density matrices.

\section{The pseudocode for physical-sector extraction}

Suppose we have a set of commuting measurement data $\{f_j\}$ that
form the column $\bm{f}$, as well as the associated outcomes $\Pi_j$
of some eigenbasis $\{\ket{0},\ket{1},\ket{2},\ldots\}$ that is
adopted to represent $\Hphys$. For some pre-chosen basis and $\alpha$
significance, the pseudocode for PSEP is presented as follows:
\begin{description}
\item [step 1.] 
Compute the measurement matrix $\bm{C}$ and sort
  $\bm{C}^{-}\bm{f}$ in descending order to obtain the ordered index
  $\bm{i}$. Then, define the ordered sequence of basis kets
  $\{\ket{n_{i_1}},\ket{n_{i_2}},\ket{n_{i_3}},\ldots\}$.
\item [step 2.] 
Set $k=0$ and $\Hsub=\Span{\ket{n_{i_1}}}$.
\item [step 3.\label{step:sloop}] 
Construct $W_{\mathrm{sub}}$ by solving the
  linear system of equations in equation~\eref{eq:dec_obs} for the $y_j$s.
\item [step 4.] 
Compute $w_\mathrm{sub}$, $\Delta^2$ and hence
  $B_\mathrm{sub}$. For typical multinomial data,
  $\Delta^2=\sum_{jk}y_jy_k(\delta_{j,k}p_j-p_jp_k)/N$.
\item [step 5.\label{step:eloop}] 
Increase $k$ by one and include
  $\ket{n_{i_k}}$ in $\Hsub$.
\item [step 6.] 
Repeat \textsc{step}~3 through 5 until
  $B_\mathrm{sub}\geq\alpha$. Finally, report $\Hphys=\Hsub$ and
  $\alpha$ and proceed to perform quantum-state tomography in
  $\Hphys$.
\end{description}

\section{Results}

\subsection{Quantum light sources}

\begin{figure}[b]
  \centering
  \includegraphics[width=0.87\columnwidth]{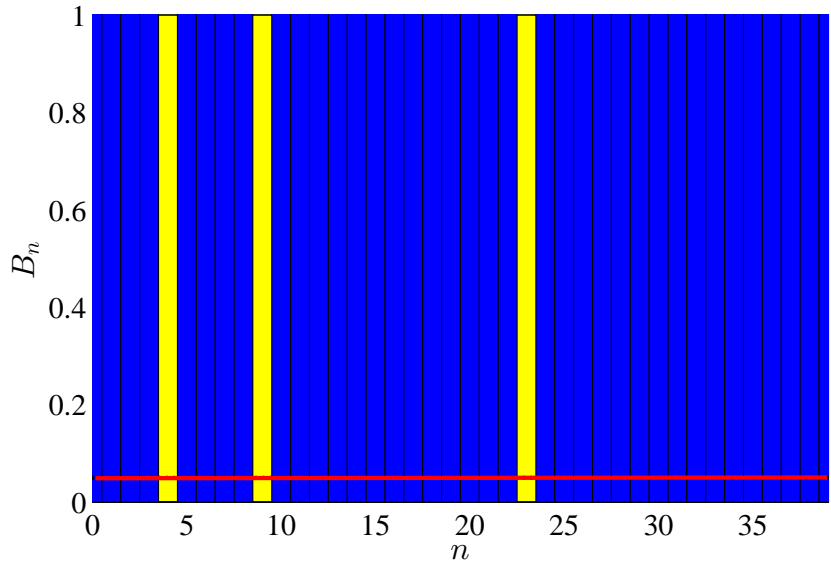}
  \caption{Physical sectors extracted with PSEP from simulated data of
    $N=10^9$ detection events for (a)~the pure state in
    \eref{eq:rho_EX} (black solid curve represents its photon-number
    distribution) and (b)~the mixed state
    $\varrho = \ket{4}\frac{1}{4}\bra{4}+ \ket{9}\frac{1}{2}\bra{9} +
    \ket{23}\frac{1}{4}\bra{23}$.  2000 random sets of 40 commuting
    measurement outcomes were used to calculate the average
    $B_{\mathrm{sub}}$ in every iterative step $k$. The (blue)
    histogram plots $B_{\mathrm{sub}}$ for the default ordering of the
    basis kets labeled with $n=0,1,2\ldots$. The physical sector
    $\Hphys$ (yellow region) is revealed after completing PSEP with
    respect to a 5\% significance level ($\alpha=0.05$) (red solid
    line).}
  \label{fig:1}
\end{figure}

To illustrate PSEP, we consider the state in \eref{eq:rho_EX} and
$\varrho = \ket{4}\frac{1}{4}\bra{4} + \ket{9}\frac{1}{2}\bra{9} +
\ket{23}\frac{1}{4}\bra{23}$. Simulated data are generated with a
random set of commuting measurement outcomes. The extracted physical
sectors are shown in figure~\ref{fig:1}.

Data statistical fluctuation may be further minimized by averaging
$B_{\mathrm{sub}}$ over many different sets of commuting
outcomes. Moreover, one can detect additional systematic errors that
are not attributed to truncation artifacts by inspecting the
corresponding histograms for errors larger than the statistical
fluctuation.

\begin{figure}[t]
  \centering
  \includegraphics[width=0.92\columnwidth]{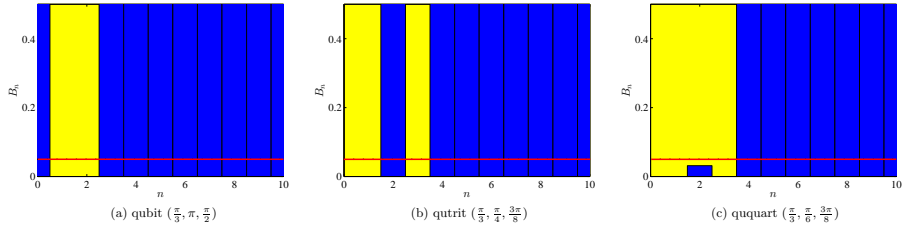}
  \caption{Schematic diagram of (a) the experimental setup to measure
    a mixture of coherent states and (b) the result of PSEP on the
    data for a mixture of two coherent states of mean photon numbers
    9.043 and 36. Panel~(a) describes coherent states from a pulsed
    laser pass through an amplitude modulator (AM), which switches
    between two values of attenuation. Neutral density (ND) filters
    further attenuate the light to the single photon level.  The
    time multiplexing detector (TMD) consists of three fiber couplers,
    delay lines and superconducting nanowire single photon detectors
    (SPD). The physical sector in panel (b) is extracted from data of
    $N=9.6\times10^6$ detection events. 5000 different sets of 60
    outcomes out of the measured 256 were used to calculate the
    average $B_{\mathrm{sub}}$ in every iterative step. Other figure
    specifications follow those of figure~\ref{fig:1}.}
  \label{fig:2}
\end{figure}

We next proceed to experimentally validate PSEP by measuring
photon-click events of a time-multiplexed detector (TMD). We use a
fiber-integrated setup to generate and measure a mixture of coherent
states, as depicted in figure~\ref{fig:2}(a). Coherent states are
produced by a pulsed diode laser with 35~ps pulses at 200~kHz and a
wavelength of 1550~nm. These pulses are then modulated with a telecom
Mach-Zehnder amplitude modulator, driven with a square-wave signal at
230~kHz. This produces pseudorandom pulse patterns with two fixed
amplitudes. After passing through fiber-attenuators, the state is
measured with an eight-bin TMD~\cite{Achilles:2003cs,Avenhaus:2010ff}
with a bin separation of 125~ns and two superconducting nanowire
detectors. We record statistics of all possible $2^8$ bin
configurations, which corresponds to a total of 256 TMD outcomes.

To characterize the TMD outcomes for the measurement, we perform
standard detector tomography, using well calibrated coherent probe
states~\cite{Rehacek:2010fk,Harder:2014tf}. The setup is similar to
the previous one, but we replace the modulator by a controllable
variable attenuator. We calibrate the attenuation with respect to a
power meter at the laser output. This allows us to produce a set of
150 probe states with a power separation of 0.2~dB.

TMD data of a statistical mixture of two coherent states are collected
and PSEP is subsequently performed on these data. The accuracy of the
extracted physical sector is ultimately sensitive to experimental
imperfections. In this case, these imperfections are minimized owing
to the state-of-the-art superconductor technology, the fruit of which
is a histogram that is as clean as it gets in an experimental
setting. Figure~\ref{fig:2}(b) provides convincing evidence of the
feasibility and practical performance of the technique, where real
data statistical fluctuation is present. This physical sector may
subsequently be taken as the objective starting point for a more
detailed investigation of the quantum signal with tools for tomography
and diagnostics.

\subsection{Finite-dimensional quantum systems}

To analyze another aspect of PSEP, in this section, we apply it to
quantum systems of finite dimensions with discrete-variable commuting
measurement outcomes. As a specific example, we consider the
arrangement in reference~\cite{Ahrens:2012aa}, which uses
single photons to encode the information simultaneously in horizontal
($H$) and vertical ($V$) polarizations, and in two spatial modes ($a$
and $b$). We define four basis states:
$|0 \rangle \equiv |H,a \rangle $, $|1 \rangle \equiv |V,a \rangle$,
$|2 \rangle \equiv |H,b \rangle$, and
$|3 \rangle \equiv |V,b \rangle$. On passing through three suitably
oriented half-wave plates at angles $\theta_{1}$, $\theta_{2}$, and
$\theta_{3}$, the state of such hybrid systems can be converted to the
pure state
$\varrho=\ket{\theta_1,\theta_2,\theta_3}\bra{\theta_1,\theta_2,\theta_3}$,
defined by
\begin{eqnarray}
  | \theta_1,\theta_2,\theta_3 \rangle  = & 
  \sin(2\theta_1)\sin(2\theta_3) \, |0 \rangle
  -\sin(2\theta_1)\cos(2\theta_3) \, |1 \rangle & \nonumber\\
  & + \cos(2\theta_1)\cos(2\theta_2) \, | 2 \rangle + 
  \cos(2\theta_1)\sin(2\theta_2) \, | 3 \rangle \,. &
   \label{eq:mpartite}
\end{eqnarray}

\begin{figure*}[t]
  \centering
  \includegraphics[width=.95\columnwidth]{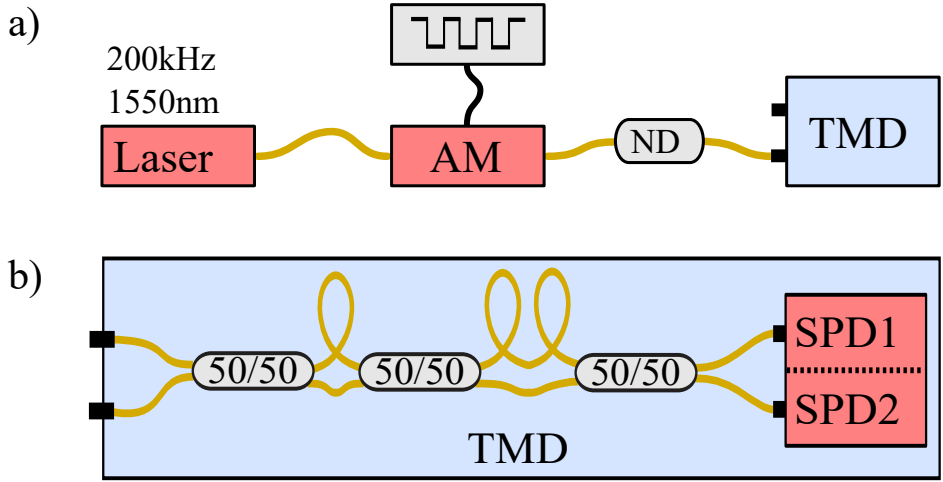}
  \caption{PSEP for hybrid quantum systems of finite dimensions that
    potentially generates either (a) a qubit state, (b) a qutrit
    state, (c) or a ququart state according to
    equation~\eref{eq:mpartite}. With $N=2.5\times10^6$ detection
    events, all three physical sectors (yellow region) are correctly
    extracted. For the ququart, the slightly higher reordered
    $B_{\Hsub}$ bar at $n=2$ (which goes to zero for larger $N$) is a
    manifestation of the favorable sensitivity of the procedure to
    specific quantum-state features, not just the overall physical
    sector. Figure specifications follow those of figure~\ref{fig:1}.}
  \label{fig:3}
\end{figure*}

Thus, by adjusting the orientation angles of the wave plates, one
could produce qubits, qutrits or ququarts from such a hybrid
source. Here, we show that PSEP can rapidly extract $\Hphys$ by
inspecting only the data measured from a set of commuting quantum
measurements. Figure~\ref{fig:3} presents the plots for a qubit,
qutrit and ququart system characterized by the different
($\theta_1,\theta_2,\theta_3$) configurations.

We have thus shown that in the typical experimental scenarios where
the measurement setup is reasonably-well calibrated, and hence
trusted, $\Hphys$ can be systematically extracted within the subspace
spanned by the measurement outcomes. This allows an observer to later
probe the details of the unknown but trusted quantum source using only
the data at hand. Notice that the relevant basis states, labeled by
$n$, form a basis for the commuting measurement on the black box. As
such, this procedure is not a bootstrapping instruction. Rather, it
systematically identifies the correct $\Hphys$ without any other
\emph{ad hoc} assertions about the source. In this way, we turn PSEP
into an efficient \emph{deterministic} dimension tester with
complexity $O(\dphys)$, as we have already learnt from
section~\ref{subsec:Hphys_not_l}.

\section{Conclusions}

We have formulated a systematic procedure to extract the physical
sector, the smallest Hilbert-subspace support, of an unknown quantum
state using only the measurement data and nothing else. This is
possible because information about the physical sector is always
entirely encoded in the data. This extraction requires only few
efficient iterative steps of the order of the physical-sector
dimension.

We demonstrated the validity and versatility of the procedure with
simulated and experimental data from quantum light sources, as well as
finite-dimensional quantum systems. The results support the clear
message that, for well-calibrated measurement devices, the physical
sector can always be systematically extracted and verified with
statistical tools, in which quantum-state tomography can be performed
accurately. No \emph{a priori} assumptions about the source, which
require additional testing, are necessary. The proposed method should
serve as the reliable solution for realistic tomography experiments in
quantum systems of complex degrees of freedom.

\ack D.~M. acknowledges support from the National Academy of Sciences
of Belarus through the program ``Convergence'', the European
Commission through the SUPERTWIN project (Contract No. 686731), and by
FAPESP (Grant No. 2014/21188-0). Y.~S.~T. acknowledges support from
the BK21 Plus Program (Grant 21A20131111123) funded by the Ministry of
Education (MOE, Korea) and National Research Foundation of Korea
(NRF). J.~{\v R} and Z.~H. acknowledge support from the Grant Agency
of the Czech Republic (Grant No. 15-03194S), and the IGA Project of
Palack{\'y} University (Grant No. PRF~2016-005). J.~T., R.~K., G.~H.,
and Ch.~S. acknowledge the European Commission through the QCumber
project (Contract No. 665148). Finally, L.~L.~S.~S. acknowledges the
Spanish MINECO (Grant FIS2015-67963-P).


\end{document}